\documentclass[prc,nofootinbib,showpacs]{revtex4}

\usepackage{graphicx}
\usepackage[usenames,dvipsnames]{color}
\usepackage{amsmath,amssymb,bbold}
\usepackage{hyperref}
\usepackage{appendix}

\def\bea {\begin{eqnarray}}
\def\eea {\end{eqnarray}}
\def\bc {\begin{center}}
\def\ec {\end{center}}
\def\nn {\nonumber}

\begin{document}

\title{Dark matter admixed neutron star properties in the light of gravitational wave observations: a two fluid approach}

\author{Arpan Das$^{1,}$\footnote{ arpan.das@ifj.edu.pl}}
\author{Tuhin Malik$^{2,}$\footnote{tuhin.malik@gmail.com}}
\author{Alekha C. Nayak$^{3,}$\footnote{ alekhacu@gmail.com}}
\affiliation{$^1$Institute of Nuclear Physics Polish Academy of Sciences, PL-31-342 Krak\'ow, Poland}
\affiliation{$^2$BITS-Pilani, Department of Physics, Hyderabad Campus, Hyderabad - 500078, India.}
\affiliation{$^3$Department of Physics, NIT Meghalaya, Bijni Complex, Laitumkhrah
Shillong-793003, Meghalaya, India.}

\date{\today} 

\begin{abstract} 
We consider the effect of density dependent dark matter on the neutron star mass, radius, and tidal deformability. Nuclear matter (normal matter) as well as the fermionic dark matter sector is considered in a mean field model. We adopt the two fluid formalism to investigate the effect of dark matter on the neutron star properties. In the two fluid picture, there is no direct interaction between the dark matter and the nuclear matter. Rather these two sectors interact only through gravitational interaction. The nuclear matter sector is described by the $\sigma-\omega-\rho$ meson interaction in the ``FSU2R" parameterization. In the dark matter sector, we use the Bayesian parameter optimization technique to fix the unknown parameters in the dark matter equation of state. In the two fluid picture, we solve the coupled Tolman-Oppenheimer-Volkoff (TOV) equations to obtain the mass and radius of dark matter admixed neutron stars (DANSs). We also estimate the effect of the density dependent dark matter sector on the tidal deformability of dark matter admixed neutron stars (DANSs).
\end{abstract}

\maketitle

\section{Introduction}
The Universe has only $\sim6\%$ visible matter and the remaining $\sim94\%$ is composed of dark matter ($~ 26\%$) and dark energy ($~68\%$).
A strong hint for the existence of dark matter (DM) came from the observations of the kinematics of self-gravitating objects such as galaxies and clusters of galaxies. Further cosmological observations indicate that DM may be baryon free and it must be a new form of matter which can interact with the rest of the standard model particles but only very weakly (for a recent review on dark matter physics see \cite{Plehn2017}). An exciting possibility of baryonic dark matter, where the dark matter is composed of quark matter nugget can be found in Ref.\cite{zhitnisky2013} and the references therein.   Although the exact properties of the dark matter particles are not known, extensive studies on dark matter models from a particle physics point of view have put strong constraints on the mass and the coupling of the dark matter particles \cite{Plehn2017}. Among various dark matter models weakly interacting massive particle (WIMP) scenario has gained favor because in this model one can get the measured relic abundance of the dark matter very naturally using only the weak interaction scale physics. Since WIMP is associated with the weak interaction scale, this allows us to study these weak scale particles in the terrestrial laboratories \cite{felix2017}.

In recent years one of the indirect methods that have gained attention is the study of the DM effects on compact stars properties. Compact stars like neutron stars (NS) provide us ample opportunities to study multidisciplinary physics such as the general theory of relativity, low energy nuclear physics, QCD under extreme conditions, etc. A neutron star is one the possible outcome of the gravitational collapse that a dying star undergoes, with a mass range between 1.4-3.0$ M_{\odot}$ ($M_{\odot}$ denotes the solar mass). In hydrostatically stable neutron stars gravitational collapse is balanced by the degeneracy pressure of neutrons which originates due to the quantum nature of neutrons. However, this neutron degeneracy pressure cannot make the star stable against the gravitational collapse if the mass of a dying star is very large. When the gravitational collapse becomes dominant as compared to the neutron degeneracy pressure, then the star cannot achieve hydrostatic equilibrium and the stellar remnant produces a black hole. Inside the neutron star, the matter density can be very large and it is generally considered to be as high as a few times nuclear saturation density ($n_0=0.16 $ fm$^{-3}$)\cite{JMLattimer2013}. Neutron stars are the only laboratories provided by Nature where such high-density matter can exist. Naturally interior of neutron stars gives us a unique opportunity to study the behavior of matter under extreme conditions. An important piece of information about the neutron star is its mass-radius relationship, which can be obtained by solving Tolman-Oppenheimer-Volkoff (TOV) equation, provided we know the nuclear equation of state (EoS) \cite{toveq}. Tolman-Oppenheimer-Volkoff (TOV) equation has been studied extensively to put constraints on the nuclear EoS \cite{JMLattimer2013}. Independent measurements of the  NS  mass and radius can also be used to constrain the nuclear EoS \cite{prakash2001,prakash2007,ozel2013}. Observations of quiescent  low-mass  X-ray binaries as well as X-ray  bursters can place some constraints on the EoS \cite{xray1,xray2,xray3,xray4,xray5,xray6,xray7}. NICER \cite{nicer1}, LOFT \cite{loft1,loft2} which deals with X-ray pulse profile observations from a hot spot on the NS surface can also put strong constraints on the neutron star mass-radius. An alternative way has opened up to understand the neutron star structure and the underlying EoS of dense matter, due to the observation of gravitational waves (GW) from a merging binary NS \cite{Abbott2017a}. In the neutron star binary merger, the tidal gravitational field of one neutron star induces a  quadrupole deformation to its companion star. During the early inspiral stage, the orbital separation between the neutron stars is large therefore, the NSs do not feel the gravitational field of their companion strongly. As the inspiral proceeds, due to the energy loss by gravitational wave radiation, the orbital separation decreases. Now the gravitational tidal field of one of the neutron stars increases in magnitude at the position of its companion and thus creating a quadrupole deformation on the companion (and vice versa). In the binary mergers when the orbital separation between the neutron stars becomes comparable with the size of neutron stars, the details of the internal structure of the binary neutron stars become important.
The induced quadrupole moment is related to the external tidal field through the tidal deformability. The tidal deformability is the response of the tidally induced quadrupole moment of an object subjected under an external tidal field. Tidal deformability is sensitive to the physical properties of a neutron star e.g., mass, radius, and tidal love number, which depend on the nuclear EoS. Hence using the observed tidal deformability of a neutron star merger, strong constraints can be put on the neutron star EoS \cite{Abbott2017}. 

It has been argued that dark matter (DM) can be captured in stars by losing their kinetic energy when the ambient dark matter particles scatter with ordinary matter in stars. Over a significant time scale, this capture of dark matter can lead to an accumulation of dark matter inside stars \cite{RajYu2018,GoldmanNussinov,GouldRoger,Kouvaris2007,KouvarisTinyakov2010,LavallazFairbairn2010}. Hence it is natural that in the absence of any segregation mechanism between normal matter and dark matter, one expects the presence of dark matter in astrophysical objects. Further dark matter can be present during the formation processes of astrophysical objects. Depending upon the nature of the dark matter particles, the type of the accreting objects and their evolution history determines the total amount of dark matter that can be trapped inside an astrophysical object. Weakly interacting massive particles (WIMP) can accumulate inside neutron stars due to elastic scattering with nucleons. Due to a large density of normal matter inside neutron stars, the collisional energy loss of the dark matter particles can be quite significant. This makes the capture of dark matter inside these dense compact objects efficient. An interesting situation appears with the accumulation of dark matter inside the compact objects, if the mass fraction of the dark matter reaches a critical value, then the dark matter can form a self-gravitating dense core supported by degeneracy pressure (Fermi degeneracy pressure). Considering the collapses of the dark core into a black hole, if it reaches the Chandrasekhar mass, a possible bound on the density, cross-section, and mass of WIMPs can be put forward \cite{GuverErkoca2014}. Further, the self-annihilating dark matter can heat old neutron stars in the galactic halo to a temperature detectable by infrared telescopes \cite{RajYu2018}. Self-annihilating dark matter can also affect the linear and angular momentum of compact objects\cite{silk2012}. This apart from the accumulation of non-self annihilating dark matter, such as asymmetric DM and mirror DM, inside the compact objects can also affect the structure of these compact objects. In this context, the dark-matter admixed neutron stars (DANSs) have been investigated extensively by many authors. In Refs, .\cite{sandin1,sandin2,mirrordarkmatter2020} authors considered the mirror dark matter in DANSs where both the normal matter (NM) and dark matter (DM) are present, but only interacting through gravity. It is important to note that due to the kinetic mixing of gauge bosons or due to unknown fields that can carry both ordinary and mirror charges, dark matter could also interact with ordinary matter through nongravitational interaction. The effect of dark matter on the neutron star properties due to nongravitational interaction also has been explored in literature. In Ref.\cite{PanotopoulosLopes2017}, the authors considered nongravitational interaction between normal matter and dark matter by considering the Higgs portal mechanism. In Ref.\cite{PanotopoulosLopes2017} Walecka type relativistic mean field model including $\sigma-\omega$ interaction for the nucleonic sector \cite{Walecka1974,SerotWalecka1986,boguta,GambhirRing1990,Ring1996} along with fermionic dark matter inside the neutron star has been considered. Using the mean field approximation effect of dark matter on the EoS and the corresponding mass-radius relation has been studied in Ref.\cite{PanotopoulosLopes2017}. Further in Ref.\cite{alekhatuhin} considering an improved Walecka type relativistic mean field model including $\sigma-\omega-\rho$ meson interaction with NL3 parameterization \cite{Ring1997,boguta2} effect of fermionic dark matter has been investigated on the neutron star EoS. It has been argued in Ref.\cite{alekhatuhin} that the presence of a high mass (as compared to the nucleon mass) dark particles inside the neutron stars soften the equation state and lower the value of tidal deformability. This implies that stiff equations of states such as the Walecka model with NL3 parameterization which are disfavor by the GW170817 observation can satisfy the tidal deformability bound, in the presence of a dark matter component (with a uniform density) which is interacting with the normal matter through a nongravitational interaction.

It is important to mention that when one considers a nongravitational interaction between normal matter and the dark matter, it is effectively a single fluid system. However one can also take an alternative and interesting approach by ignoring the nongravitational interaction between normal matter and the dark matter. This approach is also justified because the interaction between the normal matter and the dark matter can be very small, moreover, the in-medium properties of dark matter e.g. mass, self-interaction, etc., are still not known precisely. So if we ignore the nongravitational interaction between normal matter and dark matter then it is inherently a two-fluid system where normal matter and dark matter interact through gravitational interaction only. Alternative approaches have been discussed in the literature to study the two-fluid systems in the context of neutron star structure. Leung et al.in Ref \cite{leung2011}, used the general relativistic two-fluid formalism based on the master function formalism, originally developed to study the effect of nuclear superfluid on the neutron star structure \cite{masterfunction}, to study DANSs. An alternative two fluid approach has been used by Sandin et.al. to study the effect of dark matter core on the structure of neutron star \cite{sandin1,sandin2}. The same approach as discussed in Refs.\cite{sandin1,sandin2}, has been used for a broad variety of dark matter particles with various masses and interactions in neutron stars to examine the effects of dark matter on mass-radius relation of neutron stars \cite{Xiang2014}. Two-fluid formalism has been also used to study possible effects of a dark matter core on the mass-radius relation and the neutron star tidal deformability parameter for a self-interacting bosonic dark matter \cite{ellis}. 

In the present investigation, we have considered the two-fluid formalism as developed in Refs.\cite{sandin1,sandin2,Xiang2014,ellis} to study the effect of fermionic dark matter on the neutron star properties including tidal deformability, in a relativistic mean field approach. Here we consider the density dependent dark matter sector in a mean field description, i.e. relativistic mean field model (RMF) description analogous to the nuclear matter sector. So effectively here the dark matter sector is like a ``mirror" to the normal matter sector, as has been considered in Ref.\cite{Xiang2014}, but with a different equation of state as compared to the normal matter. It is important to note that in the relativistic mean field description of the normal matter all the coupling/parameters of the theory are fixed by the experimental observations. However, no such constraints are available for the density dependent dark matter sector. Therefore the parameters in the dark matter sector can be chosen in a wide range for an acceptable mass-radius relationship of neutron stars, as has been demonstrated in Ref.\cite{Xiang2014}. Naturally, with unknown parameters, we will get a large number of neutron star configurations some of which may or may not satisfy the known experimental observations. In the present investigation instead of choosing the unknown parameters in the dark matter sector arbitrarily, we use the Bayesian parameter optimization techniques to fix these unknown parameters using the knowledge of experimental and empirical information. Note that Bayesian parameter optimization techniques have been used extensively in the context of compact stars, e.g. Refs.\cite{Biswas:2020xna,Steiner:2010fz,Abbott:2018wiz,De:2018uhw,Riley:2019yda}. Due to the Bayesian parameter optimization method, we get a distribution of the parameters in the dark matter sector and we also present the results for this range of optimized parameters. The rest of the paper is organized in the following manner. 

 In Sec.~\eqref{formalism} we discuss the theoretical framework including discussions on the relativistic mean field approach to the normal matter as well as to the dark matter sector. We also include a brief discussion on Bayesian parameter optimization techniques required to fix the parameters in the dark matter sector. After the formalism part, in Sec.~\eqref{results} we present the results and demonstrate the effects of fermionic dark matter in a two fluid approach.  Finally in Sec.~\eqref{conclusion} we conclude our results with an outlook to it.

\section{Theoretical framework}
\label{formalism}
In this section, we discuss the theoretical framework. In subsection \eqref{nuclearEOS} we start with a brief discussion on the nuclear EoS in the Relativisic mean field (RMF) approach. Discussion of nuclear EoS will be followed up by a discussion on the dark matter EoS within the framework of RMF approach in subsection \eqref{darkEOS}. Further in subsection \eqref{twofluid} we discuss the theoretical framework of two fluid TOV equations along with the tidal deformability in the two fluid picture. Note that in the framework of RMF approach there are unknown parameters/couplings in the dark matter sector. We use Bayesian techniques to optimize these parameters. For completeness, we give a brief discussion on the  Bayesian parameter optimization in subsection \eqref{bayesian}.

\subsection{Nuclear matter EOS}
\label{nuclearEOS}
As stated earlier in this investigation we have considered the normal nuclear matter as well as the dark matter sector within the framework of relativistic mean field theory (RMF)
\cite{Walecka1974,SerotWalecka1986,GambhirRing1990,Ring1996}. In the RMF approach nucleons behave like quasiparticles with an effective medium dependent mass. In this model nucleon-nucleon interaction is mediated by meson exchange e.g. the exchange of $\sigma$ and $\omega$ mesons. In general, $\sigma$ mesons exchange gives rise to an attractive central force and a spin-orbit nuclear force. Further, the repulsive part of the nuclear potential can be incorporated if one also takes into account the  $\omega$-meson exchange interaction among the nucleons.  However to explain nuclear saturation properties e.g.  compressibility \cite{ChungWang} we also have to include $\rho$ meson exchange interaction between nucleons \cite{Ring1997} along with $\sigma$ and $\omega$ meson exchange. Since protons and neutrons only differ in terms of their isospin projections, the inclusion of  $\rho$ meson gives rise to a better understanding of the symmetry energy \cite{glendenning2000}. Further, for a charged neutral nuclear matter one can safely ignore the photon field.
 
The Lagrangian density of the nuclear matter sector including nucleon field, $\sigma$, $\omega$, and $\rho$ mesons and their interactions can be expressed as \cite{Tolos2017},
\begin{align}
\mathcal{L}_{\text{NM}} &=\bar{\Psi}(i \gamma_{\mu} \partial^{\mu}-M_n +g_{\sigma N} \sigma-g_{\omega N} \gamma_{\mu} \omega^{\mu}-g_{\rho N} \gamma_{\mu} \vec{I}_N. \vec{\rho}^{\mu}) \Psi \nonumber\\
& + \frac{1}{2} \partial_{\mu} \sigma \partial^{\mu} \sigma-\frac{1}{2} m_{\sigma}^{2} \sigma^{2}-\frac{\kappa}{3 !}\left(g_{\sigma N} \sigma\right)^{3}-\frac{\lambda}{4 !}\left(g_{\sigma N} \sigma\right)^{4}\nonumber \\
&-\frac{1}{4} \Omega^{\mu \nu} \Omega_{\mu \nu}+\frac{1}{2} m_{\omega}^{2} \omega_{\mu} \omega^{\mu}+\frac{\zeta}{4 !}\left(g_{\omega N}^2 \omega_{\mu} \omega^{\mu}\right)^{2} \nonumber\\
&-\frac{1}{4} \vec{R}^{\mu \nu} \vec{R}_{\mu \nu}+\frac{1}{2} m_{\rho}^{2} \vec{\rho}_{\mu} \vec{\rho}^{\mu}+\Lambda_{\omega} g_{\rho N}^{2} \vec{\rho}_{\mu} \vec{\rho}^{\mu} g_{\omega N}^{2} \omega_{\mu} \omega^{\mu} \label{Lag_m}
\end{align}
In the above equation $\Psi$ represents nucleon field, $\sigma$, $\omega_{\mu}$ and $\vec{\rho}_{\mu}$ denotes $\sigma$, $\omega$, $\rho$ 
meson field respectively. $\vec{I}_N$ denotes the isospin generators for the nucleons. $m_{\sigma}$, $m_{\omega}$ and $m_{\rho}$ are the corresponding masses of the 
mesons and $M_n$ denotes the bare nucleon mass which is different from the medium dependent nucleon mass $M_n^{\star} {}= M_n-g_{\sigma N}\sigma$. The mesonic field strength tensors are $\Omega_{\mu \nu}=\partial_{\mu} \omega_{\nu}-\partial_{\nu} \omega_{\mu}, \vec{R}_{\mu \nu}=\partial_{\mu} \vec{\rho}_{\nu}-$
$\partial_{\nu} \vec{\rho}_{\mu}$ for $\omega$ and $\rho$ mesons respectively. Nuclear manybody dynamics in encoded in the coupling constants of the above Lagrangian. The couplings of the $\sigma$ and $\omega$ mesons to the nucleons i.e. $g_{\sigma N}$ and $g_{\omega N}$ determine the energy per particle and density of the nuclear matter saturation point \cite{Li2014}. The $g_{\rho N}$ is the coupling of the isovector meson $\rho$ to the nucleon, determines the nuclear symmetry energy.
Couplings in the $\sigma$ meson self-interaction terms i.e. $\kappa$ and $\lambda$ are important for successful descriptions of nuclear matter and finite nuclei within the framework of a relativistic theory \cite{Boguta1977,Boguta1983}. These couplings give rise to a softer EoS at moderate densities and allow realistic compressibility of nuclear matter. The non negative quartic self-coupling $\zeta$ gives rise to an attractive interaction that softens the nuclear EoS at high density. Therefore this term affects the structure and maximum mass of neutron stars \cite{Bodmer1991,Mueller1996}. 
Further, the interaction between the $\omega$ meson and  $\rho$ meson denoted by the coupling $\Lambda_{\omega}$ affects the density dependence of the nuclear symmetry energy which also affects the pressure of neutron matter and the radii of neutron stars \cite{Horowitz2001}.

The covariant equation of motion or the Euler-Lagrange equation of the fundamental fields $\Psi$, $\sigma$, $\omega^{\mu}$ and $\vec{b}_{\mu}$ can be obtained using the Lagrangian densities $\mathcal{L}_{\text{NM}}$. The Euler-Lagrange equation for any  field $\varphi$ having the Lagrangian density $(\mathcal{L}_{\varphi})$ is given by,

\begin{equation}
\partial_{\mu}\left(\frac{\partial \mathcal{L}_{\varphi}}{\partial\left(\partial_{\mu}  \varphi\right)}\right)=\frac{\partial \mathcal{L}_{\varphi}}{\partial  \varphi}.
\end{equation}
The associated stress-energy tensor $T^{\mu \nu}$ is,
\begin{equation}
T^{\mu \nu}_{\varphi}=\frac{\partial \mathcal{L}_{\varphi}}{\partial\left(\partial_{\mu}  \varphi\right)} \partial^{\nu}  \varphi-g^{\mu \nu} \mathcal{L}_{\varphi}
\end{equation}
where $g_{\mu \nu}$ denotes the metric tensor and is given by $g_{\mu \nu}=\operatorname{diag}[1-1-1-1]$. The energy density $\varepsilon_{\varphi}$ and pressure $P_{\varphi}$ of the system in static case is then,
\begin{equation}
\begin{array}{c}{\varepsilon_{\varphi}=\langle T^{00}_{\varphi}\rangle}, \\ {P_{\varphi}=\frac{1}{3}\langle T^{ii}_{\varphi}\rangle}.\end{array}
\label{emtensor}
\end{equation}
Using the Euler-Lagrange equation  of the fundamental fields $\Psi$, $\sigma$, $\omega^{\mu}$ and $\vec{b}_{\mu}$ and the definition of the energy density and pressure as given in Eq.\eqref{emtensor} one can obtain the 
energy density ($\varepsilon_{\text{NM}}$) and pressure ($P_{\text{NM}}$) of the nuclear matter as (for a detailed discussion see Ref.\cite{alekhatuhin,Diener}):

\begin{align}
\varepsilon_{\text{NM}} &=\frac{1}{\pi^{2}} \int_0^{k_{p}} dk~k^2 \sqrt{k^{2}+\left(M_{n}^{*}\right)^{2}}+\frac{1}{\pi^{2}} \int_0^{k_{n}} dk~k^2 \sqrt{k^{2}+\left(M_{n}^{*}\right)^{2}}\nonumber\\
&+\frac{1}{2} m_{\sigma}^{2} \bar{\sigma}^{2}+\frac{1}{2} m_{\omega}^{2} \bar{\omega}^{2}+\frac{1}{2} m_{\rho}^{2} \bar{\rho}^{2}+\frac{\kappa}{3 !}\left(g_{\sigma N} \bar{\sigma}\right)^{3}+\frac{\lambda}{4 !}\left(g_{\sigma N} \bar{\sigma}\right)^{4}+\frac{\zeta}{8}\left(g_{\omega N} \bar{\omega}\right)^{4}+3 \Lambda_{\omega}\left(g_{\rho N} g_{\omega N} \bar{\rho} \bar{\omega}\right)^{2},
\end{align}

\begin{align}
P_{\text{NM}}=&\frac{1}{3 \pi^{2}} \int_{0}^{k_{p}} d k \frac{k^{4}}{\sqrt{k^{2}+(M_n^{*})^ 2}}+\frac{1}{3 \pi^{2}} \int_{0}^{k_{n}} d k \frac{k^{4}}{\sqrt{k^{2}+(M_n^{*})^ 2}}\nonumber\\
& -\frac{1}{2} m_{\sigma}^{2} \bar{\sigma}^{2}-\frac{\kappa}{3 !}\left(g_{\sigma N} \bar{\sigma}\right)^{3}-\frac{\lambda}{4 !}\left(g_{\sigma N} \bar{\sigma}\right)^{4}+\frac{1}{2} m_{\omega}^{2} \bar{\omega}^{2}+\frac{\zeta}{4 !}\left(g_{\omega N} \bar{\omega}\right)^{4}+\frac{1}{2} m_{\rho}^{2} \bar{\rho}^{2}+\Lambda_{\omega}\left(g_{\omega N} \bar{\omega}\right)^{2}\left(g_{\rho N} \bar{\rho}\right)^{2}
\end{align}
respectively. $\rho_n$ and $\rho_p$ are the neutron and proton number density with $k_n$ and $k_p$ are the corresponding Fermi momentum of neutron and proton, respectively. In a static and isotropic medium $\bar{\sigma}$, $\bar{\omega}$ and $\bar{\rho}$ represents the mean field values of the corresponding mesons.  
For a complete description of the EoS, the couplings have to be specified. In this investigation, we are considering the `FSU2R' parametrization on the Lagrangian as given in Eq.\eqref{Lag_m}. The numerical values of all these couplings/parameters are given in the Table \eqref{tableFSU2R} \cite{Tolos2017}.

\begin{table}
\begin{tabular}{|c|c|c|c|c|c|c|c|c|c|c|} 
 \hline
 $M_n$ & $m_{\sigma}$ & $m_{\omega}$ & $m_{\rho}$ & $g_{\sigma N}^2$&  $g_{\omega N}^2$ & $g_{\rho N}^2$ & $\kappa$ & $\lambda$ & $\zeta$ & $\Lambda_{\omega}$ \\
  (MeV) & (MeV) & (MeV) & (MeV) &  & & & (MeV)  & & & \\
 \hline
 939 & 497.479 & 782.500 & 763.000 & 107.5751 & 182.3949  & 206.4260 & 3.0911  & -0.001680 & 0.024 & 0.045 \\
 \hline
 \end{tabular}
 \caption{Parameters of the FSU2R model considered in this investigation \cite{Tolos2017}. }
 \label{tableFSU2R}
\end{table}

\subsection{Dark matter EOS}
\label{darkEOS}
Inspired by the success of the nuclear mean field approach to describe the nuclear matter, in this investigation we have modeled the dark matter sector using the mean field approach following Ref.\cite{Xiang2014}. It is important to mention that using our knowledge of mean field approach in the nuclear matter sector one can write advanced Lagrangian including various ``dark hadrons" \cite{Xiang2014}. But this comes at a price with a large number of unknown parameters. Therefore to reduce the effect of the unknown parameters in the theory we consider the dark matter Lagrangian with a single fermionic component and the self interaction of the fermionic dark matter is mediated by `dark scalar' and `dark vector' boson particles. This choice is motivated by the fact that in this minimal model attractive as well as repulsive self interaction between the fermionic dark matter particles can be obtained due to the `dark scalar' and `dark vector' mediators respectively. Therefore the dark matter Lagrangian incorporating attractive and repulsive interaction can be expressed as, 
\begin{equation}\begin{aligned}
\mathcal{L}_{DM}=& \bar{\psi}_{D}\left[\gamma_{\mu}\left(i \partial^{\mu}-g_{vd} V^{\mu}\right)-\left(M_{D}-g_{sd} \phi_D\right)\right] \psi_{D}+\frac{1}{2}\left(\partial_{\mu} \phi_D \partial^{\mu} \phi_D-m_{sd}^{2} \phi_D^{2}\right) \\
&-\frac{1}{4} V_{\mu \nu,D} V^{\mu \nu}_D+\frac{1}{2} m_{vd}^{2} V_{\mu,D} V^{\mu}_{D}.
\label{equ7}
\end{aligned}\end{equation}
Here $\psi_D$, $\phi_D$, and $V^{\mu}_D$ represent fermionic dark matter, `dark scalar meson', and `dark vector meson' respectively. $M_D$ is the `bare' mass of the fermionic dark matter. $m_{sd}$ and $m_{vd}$ are the corresponding masses of the scalar and vector `dark mesons' respectively.   
Similar to the nuclear matter equation of state one can also obtain the dark matter equation of state in the mean field approximation and is expressed as, 
\begin{align}
\varepsilon_{\text{DM}}=\frac{1}{\pi^{2}} \int_0^{k_{D}} dk~k^2 \sqrt{k^{2}+\left(M_{D}^{*}\right)^{2}}+\frac{g_{vd}^{2}}{2 m_{vd}^{2}} \rho_{D}^{2}+\frac{m_{sd}^{2}}{2 g_{sd}^{2}}\left(M_{D}-M_{D}^{*}\right)^{2},
\end{align}
\begin{align}
P_{\text{DM}}=\frac{1}{3} \frac{1}{\pi^{2}} \int_0^{k_{D}} dk \frac{k^{2}}{\sqrt{k^{2}+\left(M_{D}^{*}\right)^{2}}}+\frac{g_{vd}^{2}}{2 m_{vd}^{2}} \rho_{D}^{2}-\frac{m_{sd}^{2}}{2 g_{sd}^{2}}\left(M_{D}-M_{D}^{*}\right)^{2}.
\end{align}
Here $M_D^{*}$ is the medium dependent fermionic dark matter mass which can be given as $M_D^{*} = M_{D}-g_{sd} \phi_{D_0}$, $\phi_{D_0}$ being the mean field value of the `dark scalar meson'. $\phi_{D_0}$ can be expressed as, $\phi_{D_0}\equiv \frac{g_{sd}}{m_{sd}^2}\langle\bar{\psi}_D\psi_D\rangle$.  $\rho_D$ represents the density of the fermionic dark matter, which is associated with the mean field value of the `dark vector' meson. It is important to note that although in the dark matter Lagrangian as given in Eq.\eqref{equ7} there are four parameters, i.e. $m_{sd}, m_{vd}, g_{sd}$ and $g_{vd}$, but the dark matter energy density and pressure in the mean field approximation depends only on the ratios $g_{sd}/m_{sd}$ and $g_{vd}/m_{vd}$.  Note that this reduction of effective parameters is possible for this simple Lagrangian density in the dark matter sector. For a more complected Lagrangian density one has to deal with a large numbers of parameters in the relativistic mean field approach. For a complete description of  the dark matter equation of state we need to specify the parameters $g_{sd}/m_{sd}$ and $g_{vd}/m_{vd}$. As mentioned earlier, determination of unknown parameters in the dark matter sector using the Bayesian parameter optimization technique has been discussed in the subsequent sections.       
 
\subsection{The tidal deformability of neutron star : two fluid formalism } 
\label{twofluid}
Once we have a complete description of the nuclear as well as dark matter EoS we can study the properties of the dark matter admixed neutron stars (DANSs). To study the properties of the dark matter admixed neutron stars, we adopt a two-fluid formalism where the nuclear matter and the dark matter sectors do not interact directly, rather the interaction between these two sectors is through the gravitational interaction. The DANSs structure can be obtained by solving the hydrostatic balance equation between inward gravitational pressure and outward fermion degeneracy pressure also known as TOV equations.
The TOV equations for a two fluid system with energy density and pressure $\varepsilon_1, P_1$ and $\varepsilon_2, P_2$, where it has been assumed that the energy momentum tensors of the two fluids are separately conserved, can be expressed as (see appendix \eqref{appenA} for details)\cite{Xiang2014},

\begin{align}
 \frac{dP_1}{dr}=-(P_1+\varepsilon_1)\frac{4\pi r^3(P_1+P_2)+m(r)}{r(r-2m(r))},\label{tov1}\\
 \frac{dP_2}{dr}=-(P_2+\varepsilon_2)\frac{4\pi r^3(P_1+P_2)+m(r)}{r(r-2m(r))}\label{tov2},
 \end{align}
 where,
 \begin{align}
 \frac{dm(r)}{dr}=4\pi(\varepsilon_1(r)+\varepsilon_2(r))r^2\label{tov3}.
\end{align}
Here $\varepsilon_1,P_1$ and $\varepsilon_2,P_2$ are the energy density and pressure of the two different fluids. For this investigation one can consider $\varepsilon_1,P_1$ are associated with NM and $\varepsilon_2,P_2$ are associated with DM or vice versa.   

Apart from the mass and radius of NS, tidal deformability is also an important measurable structural property. In a coalescing binary neutron star system, during the final stages of inspiral,  each NS develops a quadrupole deformation due to the tidal gravitational field induced by the companion neutron star. The tidal deformability is the measure of the degree of deformation of a neutron star due to the tidal field of the companion star. Tidal deformability depends on the structural properties of neutron star and is sensitive to the nature of the EoS. The tidal deformability is defined as, 
\begin{equation}
\label{eq2}
\lambda = \frac{2}{3}k_2R^5,
\end{equation}
where $R$ is the radius of the NS. The value of $k_2$ is typically in 
the range $\simeq 0.05-0.15$ \cite{Hinderer2008,Hinderer2010,Postnikov2010} for NSs and depends on the stellar 
structure. This quantity can be calculated using the following expression \cite{Hinderer2008}
\bea
&& k_2 = \frac{8C^5}{5}\left(1-2C\right)^2
\left[2+2C\left(y_R-1\right)-y_R\right]\nn\\
&&\times \bigg\{2C\left(6-3 y_R+3 C(5y_R-8)\right)\nn \\
&&+4C^3\left[13-11y_R+C(3 y_R-2)+2
C^2(1+y_R)\right] \nn \\
&& ~ ~
+3(1-2C)^2\left[2-y_R+2C(y_R-1)\right]\log\left(1-2C\right)\bigg\}^{-1},
\label{eq_k2}
\eea
where $C$ $(\equiv M/R)$ is the compactness parameter of the star of
mass $M$ with radius $R$.  The quantity $y_R$ $(\equiv y(R))$ for a two fluid system can be obtained by solving
the following differential equation     
 \begin{align}
  & r \frac{d y(r)}{dr} + {y(r)}^2 + y(r) F(r) + r^2 Q(r) = 0,
   \end{align}
with,
\bea
F(r) = \frac{r-4 \pi r^3 \left( (\varepsilon_1(r)+\varepsilon_2(r)) - (P_1(r)+P_2(r)\right) }{r-2
m(r)}, \nonumber
\eea
\bea
\nonumber Q(r) &=& \frac{4 \pi r \left(5 (\varepsilon_1(r)+\varepsilon_2(r)) +9 (P_1(r)+P_2(r)) +
\frac{\varepsilon_1(r) + P_1(r)}{\partial P_1(r)/\partial
\varepsilon_1(r)}+\frac{\varepsilon_2(r) + P_2(r)}{\partial P_2(r)/\partial
\varepsilon_2(r)} - \frac{6}{4 \pi r^2}\right)}{r-2m(r)} \\
&-&  4\left[\frac{m(r) + 4 \pi r^3
(P_1(r)+P_2(r))}{r^2\left(1-2m(r)/r\right)}\right]^2, \nonumber
\eea     
along with TOV equation with proper boundary conditions \cite{tuhin2,Zhang2002} (details can be found in appendix \eqref{appenB}). One can then define the dimensionless tidal deformability as $\Lambda = \frac{2}{3}k_2 C^{-5}$. We should emphasize that for a two fluid system the evolution equation for the $y(r)$ gets modified due to presence of multiple fluid inside the neutron star. However the evolution equation of $y(r)$ outside the star remains unaltered. \\

\subsection{Bayesian parameter optimization}
\label{bayesian}
Here we discuss a framework to estimates the unknown coupling/parameters in the dark matter sector while incorporating all the necessary information at our disposal. In this context, we use the Bayesian formulation of the parameter estimation. For the parameter estimation, the use of priors is well motivated, however, we should keep in mind that priors can bias the parameter extraction as has been discussed in Ref.\cite{Wesolowski:2015fqa,Furnstahl:2015rha}. Here we are not going into a deep discussion of the use of prior because it is not our goal here. For a detailed discussion on the choice of prior and its use see Ref.\cite{Wesolowski:2015fqa}. In the Bayesian techniques, the central quantity is $p(A\mid B)$, which denotes the posterior probability distributions (PDF) of $A$ given that $B$ is true. The parameter estimation determines the posterior $p(A\mid B,I)$. $p(A\mid B, I)$ is the joint probability distribution for the full set of parameters/coefficients (in the present case couplings, etc.) denoted as  ``$A$'', given the data ``$B$'' (including their errors). $I$ represents any other information. Using the basic rules of probabilistic inference, the sum and product rules along with Bayes' theorem we can express $p(A\mid B, I)$ in terms of other probability distributions which we can calculate \cite{RTCox1946,RTCoxBook,ETJaynesBook}, e.g.,
\begin{align}
    p(A \mid B, I)=\frac{p(B \mid A, I) p(A \mid I)}{p(B \mid I)}.
\end{align}
Here $p(B\mid I)$ does not depend upon ``$A$'', therefore can be determined by normalization. $p(B \mid A, I)$ and $p(A \mid I)$ represents \textit{likelihood} and \textit{prior} respectively. In the Bayesian procedure one can identify a \textit{likelihood} function based on the available experimental data and assigning an appropriate prior, ``$A$'' can be estimated by analyzing the properties of the posterior for ``$A$''. Note that in the estimation of ``$A$'' using Bayesian techniques the choice of the prior is important. One could choose it to be uniform prior, which has been used as a baseline for many analyses. Note that in the present investigation we are estimating the parameters or couplings in the density dependent dark matter sector and we have a little prior knowledge about these parameters. Therefore we consider a uniform prior for the Bayesian analysis. Note that an important feature of Bayesian analysis is the dependence of the results on the choice of prior. However only if the data used in the analysis is not constraining then the results can depend on the choice of the prior. 

For the prior PDFs, we choose the dark matter model parameters, i.e. the ratio of scalar coupling to the scalar meson mass ($g_{sd}/m_{sd}$), the ratio of vector coupling to the vector meson mass ($g_{vd}/m_{vd}$), dark matter bare mass ($M_D$) and the ratio of the dark matter central energy density to the normal matter central energy density $(f_D\equiv \varepsilon_{DM}^c/\varepsilon_{NM}^c)$ with a uniform prior. Note that in the dark matter sector the parameters are not experimentally restricted. Therefore the choice of the dark matter parameters is guided by valuable previous investigation as has been done in Ref.\cite{Xiang2014}.

Likelihood of the model parameters $\mathbf{a}(a_1,a_2...a_j)$ with respect to the corresponding experimental data $(b_1^{\rm exp} ,b_2^{\rm exp}....b_i^{\rm exp})$ has been considered in the following form, 
\begin{align}
    p(B(b_1^{\rm exp},b_2^{\rm exp}....b_i^{\rm exp})\mid A(\mathbf{a}),I)= \Pi_i\frac{1}{\sqrt{2\pi}\sigma_i}\exp\bigg(-\frac{(b^{th}_i-b^{exp}_i)^2}{2\sigma_i^2}\bigg),
\end{align}
$\sigma_i$ denotes the width of the likelihood function. The $b_i^{\rm th}$ is the theoretically calculated observable with parameters ``$\mathbf{a}$''. The index $i$ runs from 1 to number of data. Note that sometimes models can be overfitted due to the order of the number of model parameters quite larger than the order of the number of fit data. To overcome this overfitting problem one can design a {\it likelihood function} with explicit dependence on parameter space. However, in the present analysis, the number model parameter and fit data are in the same order. Therefore, our choice of the {\it likelihood function} is not an explicit function of parameter space.

\begin{figure}[h]
\centering
\includegraphics[width=0.8\linewidth]{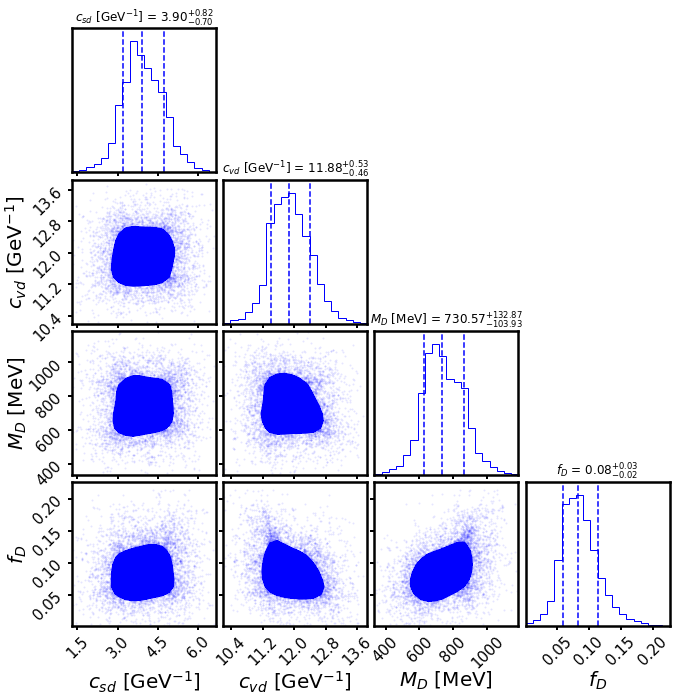}
\caption{Distribution of the dark matter sector parameters $c_{sd}$, $c_{vd}$, $M_D$ and $f_D$ as obtained by the Bayesian parameter optimization. For the Bayesian parameter optimization we have considered neutron star maximum mass, neutron star radius at 1.4 $M_{\odot}$, and tidal deformability at 1.4 $M_{\odot}$ as input data. Further for the parameters  $c_{sd}$, $c_{vd}$, $M_D$ and $f_D$  we considered a uniform prior. The posterior distribution for these parameters is given in the plot as well mentioned in the table \eqref{table1}.}
\label{fig:paradist}
\end{figure}

\section{Results}
\label{results}

To see the effects of the dark matter component on the structure of neutron stars we have to solve the TOV equations for the two fluid system as given in Eqs.\eqref{tov1},\eqref{tov2} and \eqref{tov3} once we have the full information about the equation of states (EoS) of the normal matter as well as of the dark matter. In the nuclear matter sector, various EoS has been studied extensively. As mentioned earlier in this investigation we are considering the ``FSU2R" model in the nuclear matter sector \cite{Tolos2017}. However in the dark matter sector we need to fix the following parameters, $c_{sd}\equiv g_{sd}/m_{sd}, c_{vd}\equiv g_{vd}/m_{vd}, M_D$. Note that these parameters are not fixed by any laboratory experiments. Therefore one can take some suitable choice of these parameters to fix the dark matter EoS as has been done in Ref.\cite{Xiang2014}. Apart from the dark matter EoS we also need information about the dark matter central density to solve the coupled TOV equations. We parameterize the dark matter central density as a fraction of the nuclear matter central density, ($f_D\equiv\frac{\varepsilon_{DM}^c}{\varepsilon_{NM}^c})$.
Although $f_D$ does not enter into the dark matter EoS, neutron star properties will be dependent upon the fraction of the dark matter component. The fraction of the dark matter inside neutron stars depends upon the process of the dark matter capture in compact objects. Here we also consider $f_D$ as a free parameter. We further emphasize that in the dark matter EOS, $m_{sd},g_{sd},m_{vd},g_{vd}$ are all not independent. Interestingly the dark matter EOS, as obtained in the mean field approximation only depend upon the ratios $c_{sd}\equiv g_{sd}/m_{sd}$ and $c_{vd}\equiv g_{vd}/m_{vd}$. Therefore to solve the couple TOV equation we need to fix the following parameters: $c_{sd}$, $c_{vd}$, $M_D$ and $f_D$.  One is free to choose these parameters appropriately to see their effects on the neutron star structure. It is expected that with these four unknown parameters one can come up with very exotic mass-radius plots \cite{Xiang2014}.

 In the present investigation we take an alternative approach to fix $c_{sd}$, $c_{vd}$, $M_D$ and $f_D$. Our choice of these parameters is guided by experimental and empirical information available on the properties of neutron stars. Here we use the Bayesian parameter optimization as discussed earlier to give the posterior of these four unknown parameters using the known result of the neutron star mass, neutron star radius at 1.4 $M_{\odot}$ and the tidal deformability $\Lambda$ at  1.4 $M_{\odot}$ \cite{Antoniadis,Malik,Baiotti,Tews,Annala,Abbott2018}. It is important to know that due to the absence of any prior knowledge of $c_{sd}$, $c_{vd}$, $M_D$, and $f_D$ in the dark matter sector we take a uniform prior for the analysis. After the Bayesian parameter optimization the obtained posterior for these parameters are given in Table \eqref{table1} and the distributions of these parameters are given in Fig. \eqref{fig:paradist}. For various parameters, we mention the central value of the posterior distribution along with the one sigma deviation from the central value. From Fig.\eqref{fig:paradist} it is clear that the parameters $c_{sd}$, $c_{vd}$, $M_D$ and $f_D$ are generally not correlated. 
 
\begin{figure}
    \centering
    \begin{minipage}{.48\textwidth}
        \centering
        \includegraphics[width=1.0\linewidth]{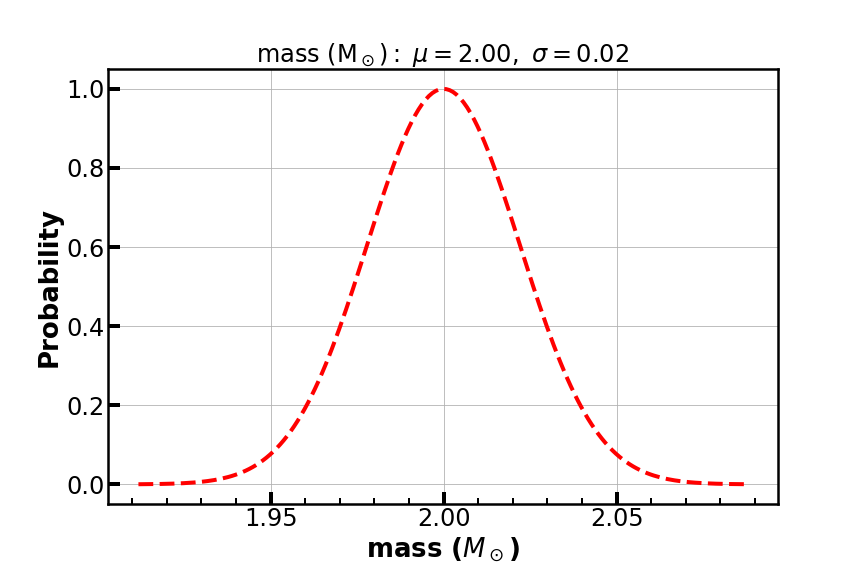}
    \end{minipage}~~
    \begin{minipage}{0.48\textwidth}
        \centering
        \includegraphics[width=1.0\linewidth]{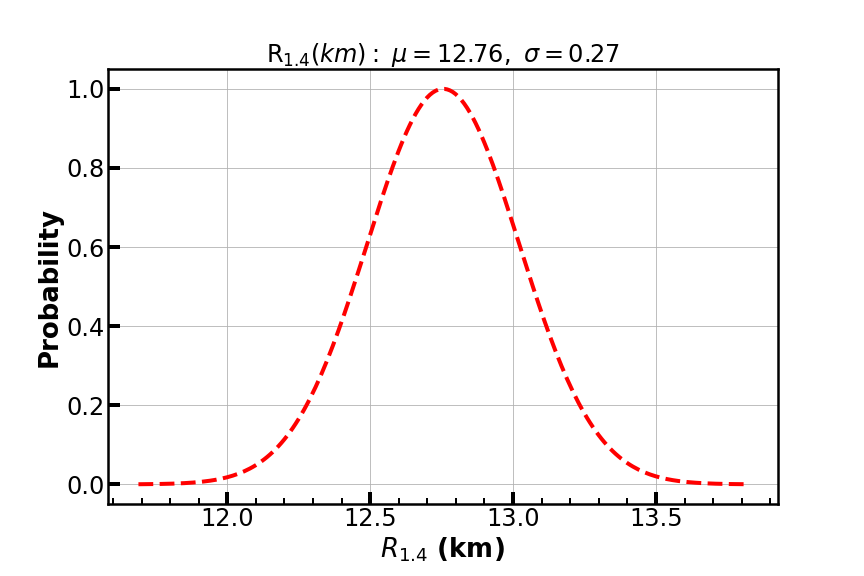}
    \end{minipage}
 \centering
 \begin{minipage}{.48\textwidth}
        \centering
        \includegraphics[width=1.0\linewidth]{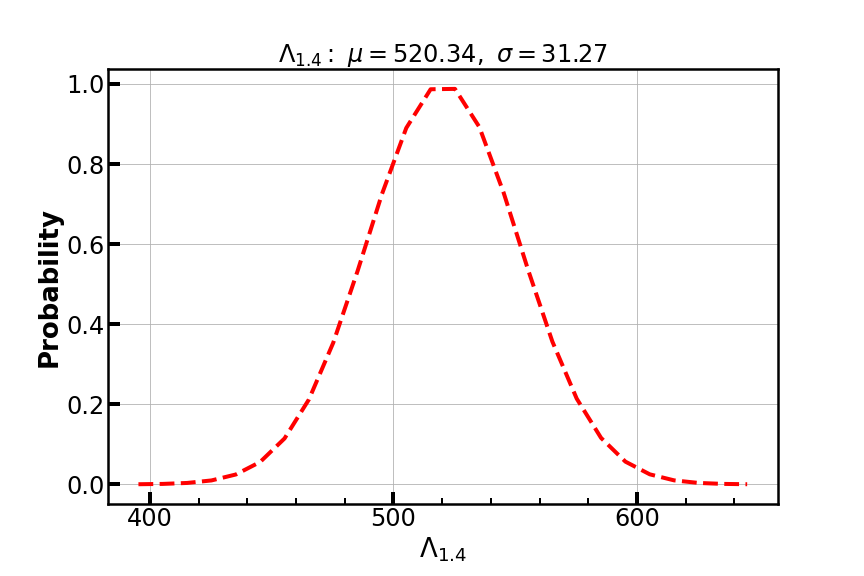}
    \end{minipage}~~
    \begin{minipage}{0.48\textwidth}
        \centering
        \includegraphics[width=1.0\linewidth]{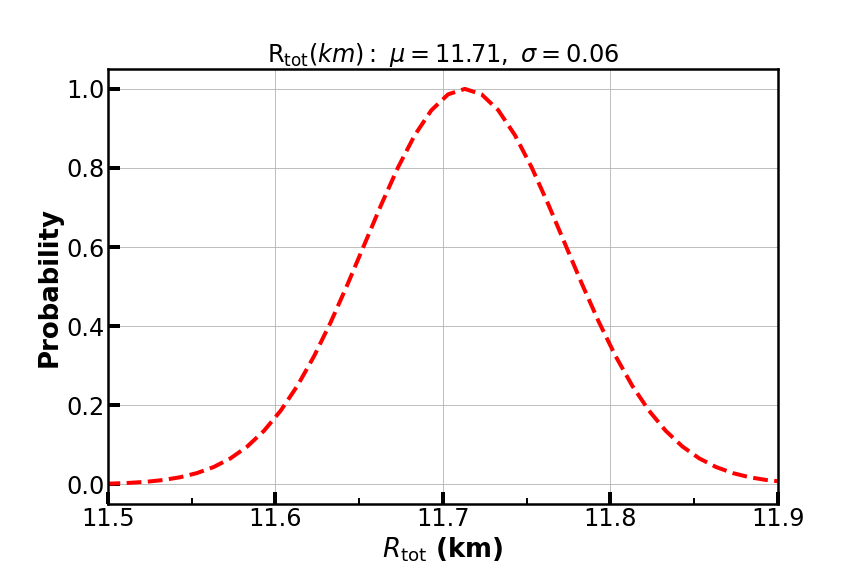}
    \end{minipage}
 \centering
 \begin{minipage}{.48\textwidth}
        \centering
        \includegraphics[width=1.0\linewidth]{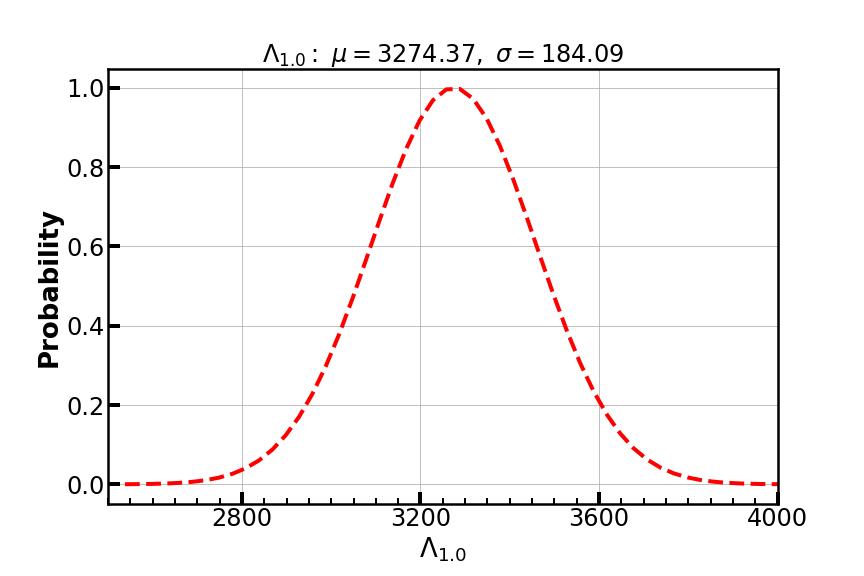}
    \end{minipage}~~
    \begin{minipage}{0.48\textwidth}
        \centering
        \includegraphics[width=1.0\linewidth]{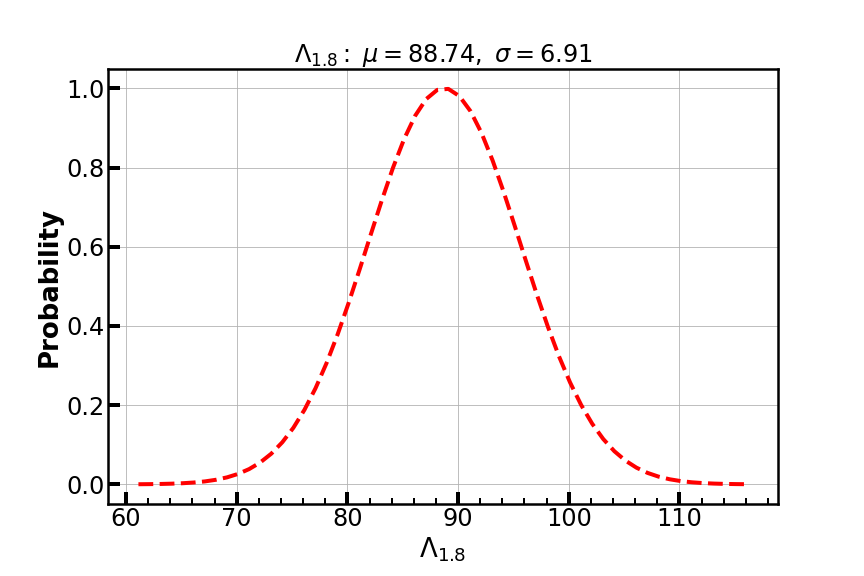}
    \end{minipage}
    \caption{In this figure we show the distribution of various physical properties of neutron star, i.e. neutron star mass, neutron star radius at 1.4 $M_{\odot}$, tidal deformability at 1.4 $M_{\odot}$, total radius of the neutron star, tidal deformability at 1.0 $M_{\odot}$ and tidal deformability at 1.8 $M_{\odot}$, for the distribution of the parameters $c_{sd}$, $c_{vd}$, $M_D$ and $f_D$ as obtained by the Bayesian parameter optimization. In these plots $\mu$ denotes the central value of the distribution and $\sigma$ denotes the standard deviation.}
\label{fig:lambdadist1}
\end{figure}

\begin{figure}
    \centering
    \begin{minipage}{.48\textwidth}
        \centering
        \includegraphics[width=1.0\linewidth]{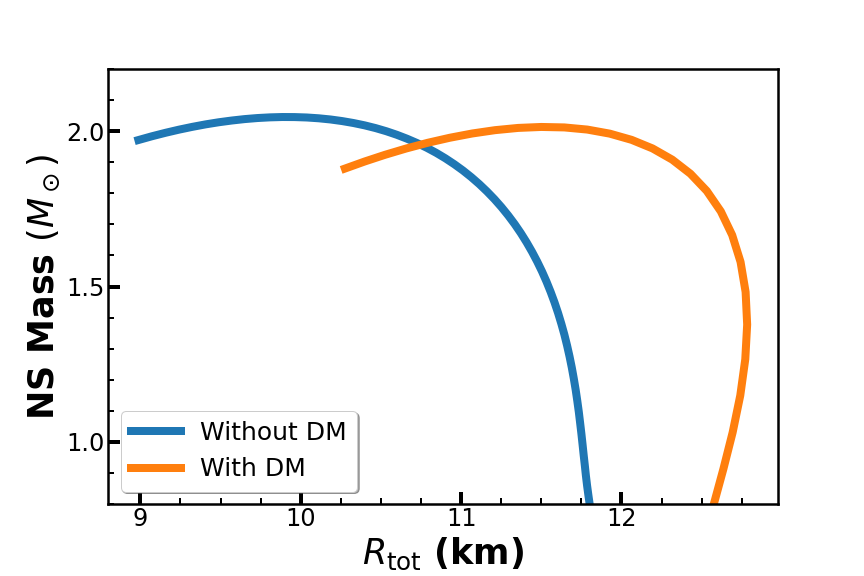}
    \end{minipage}~~
    \begin{minipage}{0.48\textwidth}
        \centering
        \includegraphics[width=1.0\linewidth]{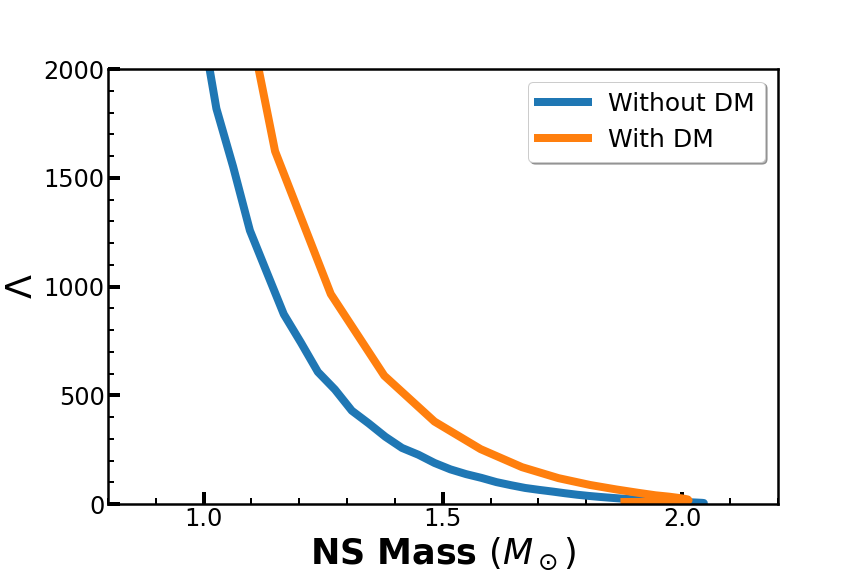}
    \end{minipage}
 \centering
        \centering
        \includegraphics[width=0.5\linewidth]{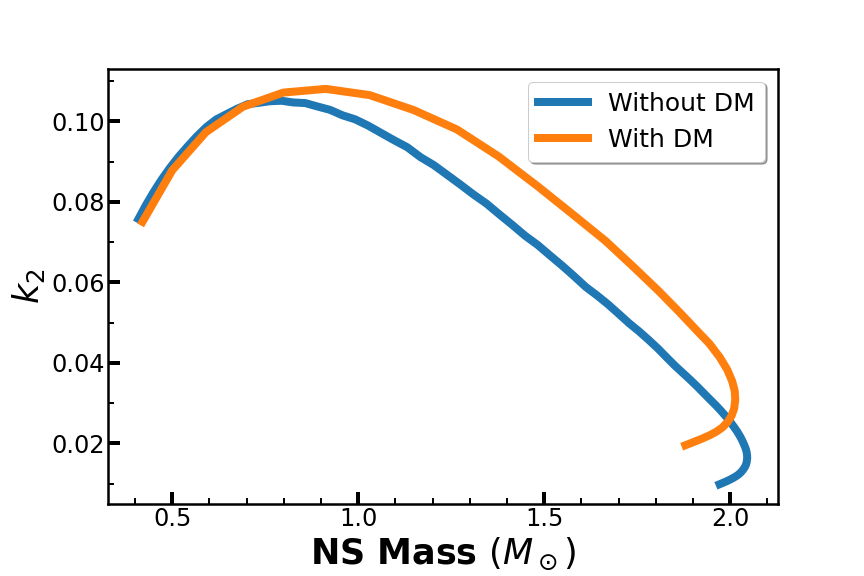}
        \caption{Mass-radius plot, tidal deformability ($\Lambda$), and $k_2$ plot with and without the dark matter component. In this plots we have considered only the central values of the parameters $c_{sd}$, $c_{vd}$, $M_D$ and $f_D$. For this configuration, we observe that the maximum mass of the neutron star decreases in the presence of the dark matter component. But the total radius corresponding to the maximum mass increases in the presence of a dark matter component. Also the tidal deformability ($\Lambda$) and $k_2$ increases in the presence of a dark matter component.}
\label{fig:nsprop}
\end{figure}

 Note that with central values of these parameters, i.e.  $c_{sd}$, $c_{vd}$, $M_D$, and $f_D$, along with the deviation from the central values there will be a large number of dark matter EoS which also corresponds to a distribution of the neutron star properties, i.e its mass, radius and the tidal deformability for a given nuclear matter EoS. Using the values of the parameters $c_{sd}$, $c_{vd}$, $M_D$ and $f_D$ as obtained from the Bayesian parameter optimization we show the distributions for the neutron star maximum mass, neutron star radius at 1.4 $M_{\odot}$ , tidal deformability at 1.4 $M_{\odot}$ along with other physical properties, e.g. neutron star radius corresponding to maximum mass $R_{tot}$, tidal deformability at 1.0 $M_{\odot}$ and tidal deformability at 1.8 $M_{\odot}$ in Fig.\eqref{fig:lambdadist1}. Further to demonstrate the effect of dark matter on the neutron star mass, radius, and tidal deformability we also provide the mass-radius and tidal deformability plots in the presence, as well as in the absence of dark matter component in Fig.\eqref{fig:nsprop}. In Fig.\eqref{fig:nsprop} for the dark matter sector we have only considered the central values of the parameters $c_{sd}$, $c_{vd}$, $M_D$ and $f_D$ for convenience.  For such configuration, from Fig.\eqref{fig:nsprop} it can be observed that with dark matter the maximum mass of the neutron star decreases, and the corresponding radius increases. Note that tidal deformability and love number scales with radius. Therefore with an increasing radius, the tidal love number and the tidal deformability also increases. This is a very distinct result as compared to our previous investigation in Ref.\cite{alekhatuhin}, where we had considered the effect of the heavy fermionic dark matter on the neutron star mass, radius, and tidal deformability. However in Ref.\cite{alekhatuhin} we considered the normal matter and the dark matter in a single fluid picture, where the dark matter interacts with the normal matter through a Higgs portal interaction.  In such a single fluid picture the maximum mass as well the corresponding radius decreases in the presence of a dark matter component.  This may be an important difference between the single fluid picture and the two fluid picture as investigated here.

\begin{table}[]
\begin{tabular}{|ccc|ccc|}
\hline
\multicolumn{3}{|c|}{Data}     & \multicolumn{3}{c|}{Parameters}                   \\ \hline
                                  & value             &   &       & Flat Prior & Posterior                     \\
NS maximum mass ($ M_{\odot}$)     & 2.01 $\pm$ 0.04  \cite{Antoniadis} &  & $c_{sd}$ (Gev$^{-1}$) & $1 -7$    & $3.90_{-0.70}^{+0.82}$       \\
$R_{1.4}$ (Km.)     & 11.3 $\pm$ 1.0 \cite{Malik,Baiotti,Tews,Annala} &   & $c_{vd}$ (Gev$^{-1}$)  & $10-14$    & $11.88_{-0.46}^{+0.53}$      \\
$\Lambda_{1.4}$     & 70-580  \cite{Abbott2018} &   & $M_D$ (MeV) & $200-1200$  & $730.57_{-103.93}^{+132.87}$ \\
 &    &   & $f_D$  & $0.0-0.25$ & $0.08_{-0.02}^{+0.03}$       \\ \hline
\end{tabular}
\caption{Table of information considered for the estimation of the unknown parameters in the dark matter sector along with the optimized parameters using the Bayesian parameter optimization. Here $\pm$ indicates the error in the values of various quantities.}
\label{table1}
\end{table}

 \section{Conclusions}
 \label{conclusion}
 In the present investigation, we consider the effect of density dependent dark matter on the neutron star properties in a two fluid framework. We modeled the density dependent dark matter within the framework of mean field theory keeping in mind the success of the nuclear mean field theory for the compact objects. In the two fluid approach neutron star properties depend on the nuclear matter equation of state as well as the dark matter equation of state. Contrary to the nuclear matter, parameters in the density dependent dark matter sector are not fixed. Using the Bayesian parameter optimization technique we fix the parameters in the dark matter sector. For a fixed nuclear equation of state, we obtain a distribution of parameters in the dark matter sector. The parameters in the dark matter sector also give rise to a distribution of various configurations of the dark matter admixed neutron stars (DANSs). Interestingly there are some configurations of DANSs where the maximum mass of the neutron star decrease in the presence of dark matter but the corresponding radius increases. For such configurations tidal deformability and the love number increase in the presence of dark matter components. 
 In Ref.\cite{alekhatuhin} earlier we obtained that in a single fluid picture where the dark matter interacts with the normal matter through the Higgs portal mechanism neutron star maximum mass and the tidal deformability decreases in the presence of dark matter. Therefore the effect of the density dependent dark matter on the neutron star properties in a two fluid approach can be significantly different as compared to the results obtained in a single fluid picture in Ref.\cite{alekhatuhin}.
  Note that in the present investigation we modeled the dark matter sector using a specific mean field model incorporating attractive and repulsive interactions. Therefore naturally the results presented here are not model independent. A model independent analysis of density dependent dark matter on the neutron star properties will be discussed in future investigations.

     \section*{Acknowledgements}
      T.M. would like to thank Prof. Hiranmaya Mishra for his support and acknowledges the hospitality provided by Physical Research Laboratory (PRL), Ahmedabad, India, during his visit, where the work was initiated. T.M would like to acknowledge the DAE-BRNS grant  No.37(3)/14/12/2018-BRNS for support. The work of A.D. is supported by the Polish National Science Center Grants No. 2018/30/E/ ST2/00432. A.D. would also like to thank Wei-Zhou Jiang for very useful clarification about the parameter fixing in the dark matter sector in their work. 
      \appendix
      \section{Two fluid TOV}
      \label{appenA}
      We start with the background Einstein equation in the presence of two distinct fluids with conserved energy momentum tensors $\overline{T}_{\mu\nu}^{(1)}$ and $\overline{T}_{\mu\nu}^{(2)}$ respectively, 
\begin{align}
 \overline{G}_{\mu\nu} & =8\pi\overline{T}_{\mu\nu}= 
 8\pi(\overline{T}_{\mu\nu}^{(1)}+\overline{T}_{\mu\nu}^{(2)}),\nonumber\\
 & = \text{diag}(e^{\alpha}\varepsilon,e^{\beta}P,r^2P
 ,r^2\sin^2\theta P),
 \end{align}
where we have identified the total energy density $(\varepsilon)$ and the total pressure $(P)$ of the two fluid system as, 
\begin{align}
 & \varepsilon(r)=\varepsilon_1(r)+\varepsilon_2(r),\\
 & P(r) =P_1(r)+P_2(r),
\end{align}
respectively. The background spherically symmetric static metric and its inverse in the spherical polar coordinate system $x^{\mu}\equiv(t,r,\theta,\phi)$, can be given as, 
\begin{align}
  & g^{(0)}_{\mu\nu}= \text{diag}\left(-e^{\alpha(r)},e^{\beta(r)},r^2,r^2\sin^2\theta\right),\\
  & g^{(0)\mu\nu}= \text{diag}\left(-e^{-\alpha(r)},e^{-\beta(r)},1/r^2,1/r^2\sin^2\theta\right).
  \end{align}
The ``$tt$" component of the background Einstein equation gives us,  
\begin{align}
 & \overline{G}_{tt}=8\pi \overline{T}_{tt}\nonumber\\
 \implies & r\beta^{\prime}=1-e^{\beta}+8\pi r^2(\varepsilon_1+\varepsilon_2)e^{\beta}.
 \label{equA6}
\end{align}
Similarly the ``$rr$" component of the background Einstein equation gives us,
\begin{align}
 & \overline{G}_{rr}=8\pi \overline{T}_{rr}\nonumber\\
 \implies & r\alpha^{\prime}=8\pi r^2(P_1+P_2)e^{\beta}+e^{\beta}-1.
 \label{equA7}
\end{align}
Eq.\eqref{equA6} can be recasted as,
\begin{align}
   & r\beta^{\prime}  =1-e^{\beta}+8\pi r^2(\varepsilon_1+\varepsilon_2)e^{\beta}\nonumber\\
  \implies & -\frac{d}{dr}(r(e^{-\beta}-1))=8\pi(\varepsilon_1+\varepsilon_2)r^2\nonumber\\
  \implies & e^{-\beta} =1-\frac{2m(r)}{r},
 \label{equA8}
\end{align}
where the mass function $m(r)$ for the two fluid system is defined as,
\begin{align}
 m(r)\equiv 4\pi \int_0^r(\varepsilon_1(r^{\prime})+\varepsilon_2(r^{\prime}))r^{\prime~2}dr^{\prime}.
\end{align}
In a similar way using Eq.\eqref{equA8}, Eq.\eqref{equA7} can be written as,
\begin{align}
 & r\alpha^{\prime}=8\pi r^2(P_1+P_2)e^{\beta}+e^{\beta}-1\nonumber\\
\implies & \frac{d\alpha}{dr}= \frac{8\pi r^3(P_1+P_2)+2m(r)}{r(r-2m(r))}.
\label{equA10}
\end{align}
Further, the conservation of the energy momentum tensors of these two fluids imply,
\begin{align}
 \nabla_{\mu}T_{(1)}^{\mu\nu}=0,\label{equA11}\\
 \nabla_{\mu}T_{(2)}^{\mu\nu}=0.
 \label{equA12}
 \end{align}
Similar to the single fluid case, in a two fluid scenario conservation of energy-momentum tensor as given by Eqs.\eqref{equA11} and \eqref{equA12} boil down to,
\begin{align}
 \frac{dP_1}{dr}=-\frac{1}{2}(P_1+\varepsilon_1)\frac{d\alpha}{dr}\label{equA13},\\
  \frac{dP_2}{dr}=-\frac{1}{2}(P_2+\varepsilon_2)\frac{d\alpha}{dr}\label{equA14}.
\end{align}
Using Eq.\eqref{equA10} back into Eqs.\eqref{equA13} and
Eq.\eqref{equA14} we obtain,
\begin{align}
 \frac{dP_1}{dr}=-(P_1+\varepsilon_1)\frac{4\pi r^3(P_1+P_2)+m(r)}{r(r-2m(r))},\label{equA15}\\
 \frac{dP_2}{dr}=-(P_2+\varepsilon_2)\frac{4\pi r^3(P_1+P_2)+m(r)}{r(r-2m(r))}\label{equA16},
 \end{align}
 with,
 \begin{align}
 \frac{dm(r)}{dr}=4\pi(\varepsilon_1(r)+\varepsilon_2(r))r^2\label{equA17}.
\end{align}
Eq.\eqref{equA15}, Eq.\eqref{equA16} along with Eq.\eqref{equA17} constitutes the TOV equations for a two fluid system. One also gets the standard TOV equation for single fluid case by taking correct limit of the two fluid TOV equations. For an alternative approach to derive the two fluid TOV equations using the techniques of Lagrange multipliers, see Ref.\cite{Xiang2014}. 

\section{Tidal deformability in a two fluid scenario}
\label{appenB}
Keeping only terms up to first order in metric perturbation, the perturbed metric and the inverse metric can be expressed as \cite{Hinderer2008},
\begin{align}
 & g_{\mu\nu}=g^{(0)}_{\mu\nu}+h_{\mu\nu},\nonumber\\
 & g^{\mu\nu}=g^{(0)\mu\nu}-h^{\mu\nu},
\end{align}
 respectively. Here $g^{(0)}_{\mu\nu}$ is the background spherically static metric and $h_{\mu\nu}$ is the associated metric perturbation. Note that $h_{\mu\nu}$ has correct tensorial properties in the background metric. Otherway stated space-time indices of $h_{\mu\nu}$ can be raised by $g^{(0)\mu\nu}$. Hence, 
 \begin{align}
  h^{\mu\nu}=g^{(0)\mu\lambda}g^{(0)\nu\sigma}h_{\lambda\sigma}.
 \end{align}
 The spherically symmetric static background metric and it's inverse can be expressed as,
 \begin{align}
  & g^{(0)}_{\mu\nu}= \text{diag}\left(-e^{\alpha(r)},e^{\beta(r)},r^2,r^2\sin^2\theta\right),\\
  & g^{(0)\mu\nu}= \text{diag}\left(-e^{-\alpha(r)},e^{-\beta(r)},1/r^2,1/r^2\sin^2\theta\right),
  \end{align}
respectively. For $l=2,m=0$, static, even-parity metric perturbations ($h_{\mu\nu}$ and $h^{\mu\nu}$) in the Regge-Wheeler gauge can be expressed as (for a detailed discussion see \cite{Hinderer2008}) , 
 \begin{align}
 & h_{\mu\nu}=\text{diag}\left(-e^{\alpha(r)}H(r)\text{Y}_{20}(\theta,\phi),-e^{\beta(r)}H(r)\text{Y}_{20}(\theta,\phi),
 r^2K(r)\text{Y}_{20}(\theta,\phi),r^2\sin^2\theta K(r)\text{Y}_{20}(\theta,\phi)\right),\\
  & h^{\mu\nu}=\text{diag}\left(-e^{-\alpha(r)}H(r)\text{Y}_{20}(\theta,\phi),-e^{-\beta(r)}H(r)\text{Y}_{20}(\theta,\phi),
 \frac{1}{r^2}K(r)\text{Y}_{20}(\theta,\phi),\frac{1}{r^2\sin^2\theta} K(r)\text{Y}_{20}(\theta,\phi)\right).
\end{align}

Using the perturbed metric and energy-momentum tensor, the first order Einstein equations can be written as, 
\begin{align}
 & \delta G^t_{~t}=8\pi\delta T^t_{~t}=-8\pi\delta\varepsilon,\label{equB7}\\
 & \delta G^r_{~r}=8\pi\delta T^r_{~r}=8\pi\delta P,\label{equB8}\\
 & \delta G^{\theta}_{~\theta}=8\pi\delta T^{\theta}_{~\theta}=8\pi\delta P,\label{equB9}\\
 & \delta G^{\phi}_{~\phi}=8\pi\delta T^{\phi}_{~\phi}=8\pi\delta P,\label{equB10}\\
 & \delta G^r_{~\theta}= 0.
\end{align}

Here the components of the perturbed Einstein tensor are, 
\begin{align}
 \delta G^{t}_{~t}= & \frac{e^{-\beta}}{2r^2}\text{Y}_{20}(\theta,\phi)\bigg[r\bigg\{2H^{\prime}+K^{\prime}(6-r\beta^{\prime})+2rK^{\prime\prime}\bigg\}
+H\left(2+6e^{\beta}-2r\beta^{\prime}\right)-4Ke^{\beta}\bigg],\label{equB12}\\
\delta G^{r}_{~r}
 = & \frac{\text{Y}_{20}(\theta,\phi)}{2r^2}e^{-\beta}\bigg[-4Ke^{\beta}+H(2+2r\alpha^{\prime}-6e^{\beta})
 +r\bigg\{2H^{\prime}+K^{\prime}(2+r\alpha^{\prime})\bigg\}\bigg]\label{equB13},\\
 \delta G^{\theta}_{~\theta}=& \frac{\text{Y}_{20}(\theta,\phi)}{4r}e^{-\beta}\bigg[2H(\alpha^{\prime}-\beta^{\prime})
 +Hr\alpha^{\prime}(\alpha^{\prime}-\beta^{\prime})+H^{\prime}(4+3r\alpha^{\prime}-r\beta^{\prime})\nonumber\\
& ~~~~~~~~~~~~~~~~~~~~~~~~~+K^{\prime}(4+r\alpha^{\prime}-r\beta^{\prime})+2rH^{\prime\prime}+2rK^{\prime\prime}+2rH\alpha^{\prime\prime}\bigg],\label{equB14}\\
\delta G^r_{~\theta}
 & = -\frac{1}{2}e^{-\beta}\frac{\partial Y_{20}(\theta,\phi)}{\partial\theta}\bigg[\alpha^{\prime}H(r)+H^{\prime}+K^{\prime}\bigg].
 \end{align}
Using Eq.\eqref{equB9}, Eq.\eqref{equB10} and Eq.\eqref{equB14} perturbation in total pressure of the two fluid system can be expressed as, 
\begin{align}
 \delta P = & \frac{\delta G^{\theta}_{~\theta}+\delta G^{\phi}_{~\phi}}{16\pi}=\frac{\delta G^{\theta}_{~\theta}}{8\pi}\nonumber\\
 = &  \frac{\text{Y}_{20}(\theta,\phi)}{32\pi r}e^{-\beta}\bigg[2H(\alpha^{\prime}-\beta^{\prime})
 +Hr\alpha^{\prime}(\alpha^{\prime}-\beta^{\prime})+H^{\prime}(4+3r\alpha^{\prime}-r\beta^{\prime})\nonumber\\
& ~~~~~~~~~~~~~~~~~~~~~~~~~~~~+K^{\prime}(4+r\alpha^{\prime}-r\beta^{\prime})+2rH^{\prime\prime}+2rK^{\prime\prime}+2rH\alpha^{\prime\prime}\bigg].
\label{equB16}
\end{align}
 The $(r,\theta)$ component of the perturbed Einstein equation allows us to write,
\begin{align}
& K^{\prime\prime}=-H^{\prime\prime}-\alpha^{\prime}H^{\prime}-\alpha^{\prime\prime}H.
\label{equB17}
 \end{align}
Using the ``$tt$" and ``$rr$" component of the perturbed Einstein tensor as given in Eq.\eqref{equB7} and Eq.\eqref{equB8} we obtain,  
\begin{align}
  & \delta G^t_{~t}-\delta G^r_{~r}  =8\pi(\delta T^0_{~0}-\delta T^r_{~r}),\nonumber\\
\implies & \delta G^t_{~t}-\delta G^r_{~r} +8\pi\bigg(1+\frac{d\varepsilon}{dP}\bigg)\delta P=0.
\label{equB18}
 \end{align}
 Here we have expressed $\delta\varepsilon=(d\varepsilon/dP)\delta P$.
Further using Eq.\eqref{equB12}, Eq.\eqref{equB13}, Eq.\eqref{equB16} and Eq.\eqref{equB17} allows us to recast Eq.\eqref{equB18} as, 

\begin{align}
 H^{\prime\prime}+C_1H^{\prime}+C_0H=0, 
 \label{equB19}
 \end{align}
where the coefficients of $H^{\prime}$ and $H$ are,
\begin{align}
 C_1 & = \frac{2}{r}+\frac{\alpha^{\prime}-\beta^{\prime}}{2},
 \label{equB20}
\end{align}
and, 
\begin{align}
 C_0
 & = \bigg[-\frac{6}{r^2}e^{\beta}+\bigg(\alpha^{\prime\prime}
 -\frac{\alpha^{\prime}\beta^{\prime}}{2}\bigg)-\frac{1}{2}\alpha^{\prime~2}
 +\frac{7}{2}\frac{\alpha^{\prime}}{r}+\frac{1}{2r}\frac{d\varepsilon}{dP}\alpha^{\prime}
 +\frac{3}{2}\frac{\beta^{\prime}}{r}+\frac{1}{2r}\frac{d\varepsilon}{dP}\beta^{\prime}\bigg],
\end{align}
respectively. Till this point the derivation of the evolution equation of $H(r)$ is general and does not assume any specific form of matter energy momentum tensor. Recall for a two fluid system, 
\begin{align}
 r\alpha^{\prime}=8\pi(P_1+P_2)r^2e^{\beta}+e^{\beta}-1,\\
 r\beta^{\prime}=1-e^{\beta}+8\pi(\varepsilon_1+\varepsilon_2)r^2e^{\beta}.
 \end{align}
Therefore $C_1$ as given in Eq.\eqref{equB20} can be expressed as,
\begin{align}
 C_1 & =\frac{2}{r}+4\pi r e^{\beta}((P_1+P_2)-(\varepsilon_1+\varepsilon_2))
 +\frac{e^{\beta}-1}{r}\nonumber\\
 & =\frac{2}{r}+e^{\beta}\bigg(\frac{2m}{r^2}+4\pi r((P_1+P_2)-(\varepsilon_1+\varepsilon_2))\bigg).
\end{align}
Recall, 
\begin{align}
 C_0=-\frac{6}{r^2}e^{\beta}+\bigg(\alpha^{\prime\prime}
 -\frac{\alpha^{\prime}\beta^{\prime}}{2}\bigg)-\frac{\alpha^{\prime~2}}{2}
 +\frac{7}{2}\bigg(\frac{\alpha^{\prime}}{r}\bigg)+\frac{1}{2r}\frac{d\varepsilon}{dP}
 \alpha^{\prime}+\frac{3}{2}\frac{\beta^{\prime}}{r}+\frac{1}{2r}\frac{d\varepsilon}{dP}\beta^{\prime}.
 \label{equB25}
\end{align}
Further using the background Einstein equation for $(\theta\theta)$ component allows us to write,
\begin{align}
 & \overline{G}_{\theta\theta}=8\pi(P_1+P_2)r^2\nonumber\\
 \implies&  \alpha^{\prime\prime}-\frac{\alpha^{\prime}\beta^{\prime}}{2}
 =16\pi(P_1+P_2)e^{\beta}-\frac{\alpha^{\prime~2}}{2}+\frac{\beta^{\prime}}{r}
 -\frac{\alpha^{\prime}}{r}.
 \label{equB26}
\end{align}
Using Eq.\eqref{equB26}, $C_0$ as given in Eq.\eqref{equB25} can be simplified to, 
\begin{align}
 C_0 
 & =-\frac{6}{r^2}e^{\beta}+4\pi e^{\beta}\bigg((P_1+P_2)+(\varepsilon_1+\varepsilon_2)\bigg)
 \frac{d\varepsilon}{dP}+4\pi e^{\beta}\bigg(5(\varepsilon_1+\varepsilon_2)+9(P_1+P_2)\bigg)
 -\alpha^{\prime~2}.
\end{align}
For two fluid system assuming both fluids have a barotropic equation of state we get,
\begin{align}
 \frac{d\varepsilon}{dP}
 = \bigg(\frac{d\varepsilon_1}{dP_1}\bigg)\frac{\delta P_1}{\delta P_1+\delta P_2}+
 \bigg(\frac{d\varepsilon_2}{dP_2}\bigg)\frac{\delta P_2}{\delta P_1+\delta P_2}.
\end{align}
Further using two fluid TOV equations (Eqs.\eqref{equA15} and \eqref{equA16}) we can write, 
\begin{align}
 \frac{\delta P_1}{\delta P_1+\delta P_2}
 = \frac{P_1+\varepsilon_1}{(P_1+P_2)+(\varepsilon_1+\varepsilon_2)},
\end{align}
and
\begin{align}
 \frac{\delta P_2}{\delta P_1+\delta P_2} 
 = \frac{P_2+\epsilon_2}{(P_1+P_2)+(\varepsilon_1+\varepsilon_2)}.
\end{align}
Therefore for two fluid case, 
\begin{align}
 C_0=-6\frac{e^{\beta}}{r^2}+4\pi e^{\beta}\bigg[5\varepsilon_1+9P_1+(P_1+\varepsilon_1)
 \bigg(\frac{d\varepsilon_1}{dP_1}\bigg)+5\varepsilon_2+9P_2+(P_2+\varepsilon_2)
 \bigg(\frac{d\varepsilon_2}{dP_2}\bigg)\bigg]-\alpha^{\prime~2}.
\end{align}
Now let us introduce the dimensionless variable $y\equiv \frac{rH^{\prime}}{H}$. In terms of $y(r)$ the evolution equation of $H(r)$ as given in Eq.\eqref{equB19} can be expressed as, 

 \begin{align}
r \frac{d y(r)}{dr} + {y(r)}^2 + y(r) F(r) + r^2 Q(r) = 0,
   \end{align}
with,
\bea
F(r) = \frac{r-4 \pi r^3 \left( (\varepsilon_1(r)+\varepsilon_2(r)) - (P_1(r)+P_2(r)\right) }{r-2
m(r)}, \nonumber
\eea
and
\bea
\nonumber Q(r) &=& \frac{4 \pi r \left(5 (\varepsilon_1(r)+\varepsilon_2(r)) +9 (P_1(r)+P_2(r)) +
\frac{\varepsilon_1(r) + P_1(r)}{\partial P_1(r)/\partial
\varepsilon_1(r)}+\frac{\varepsilon_2(r) + P_2(r)}{\partial P_2(r)/\partial
\varepsilon_2(r)} - \frac{6}{4 \pi r^2}\right)}{r-2m(r)} \\
&-&  4\left[\frac{m(r) + 4 \pi r^3
(P_1(r)+P_2(r))}{r^2\left(1-2m(r)/r\right)}\right]^2, \nonumber
\eea

For the two fluid system only the Einstein equation inside the matter changes but the evolution equation of $y$ outside matter remains unaltered (see Ref.~\cite{Hinderer2008} for details). The boundary conditions for $y(r)$ can be obtained by requiring regularity of $y(r)$ at $r=0$. By considering only the leading order contribution to $y(r)$ at $r=0$ one can easily obtain the boundary condition which is also same for a single fluid system (see Ref.~\cite{Hinderer2008}). This this because at $r=0$ in the leading order, matter part does not contribute to $y(r)$. Therefore even for multiple fluids $y(r=0)=2$.

\end{document}